\documentclass[%
 reprint,
 amsmath,amssymb,
 aps,prl
]{revtex4-2}
\usepackage{color}
\usepackage{graphicx}
\usepackage{dcolumn}
\usepackage{bm}

\newcommand{\com}[1]{\textcolor{red}{#1}}
\usepackage[english]{babel}

\begin{document}

\selectlanguage{english}


\title{Power-law intermittency in the gradient-induced self-propulsion of colloidal swimmers}

\author{Nick Oikonomeas-Koppasis$^1$, Stefania Ketzetzi$^{2,*}$, Daniela J. Kraft$^2$, Peter Schall$^1$}
\affiliation{$^1$Institute of Physics, University of Amsterdam, Science Park 904, P.O. Box 94485, 1090 GL,  Amsterdam, The Netherlands}

\affiliation{$^2$Soft Matter Physics, Huygens-Kamerlingh Onnes Laboratory, Leiden University,P.O. Box 9504, 2300 RA Leiden, The Netherlands}

\affiliation{$^*$Current address: Laboratory for Soft Materials and Interfaces, Department of Materials, ETH Zürich, Zürich, Switzerland}

\date{\today}

\begin{abstract}
Active colloidal microswimmers serve as archetypical active fluid systems, and as models for biological swimmers. Here, by studying in detail their velocity traces, we find robust power-law intermittency with system-dependent exponential cut off. We model the motion by an interplay of the field gradient-dependent active force and the locally fluctuating hydrodynamic drag, set by the wetting properties of the substrate. The model closely describes the velocity distributions of two disparate swimmer systems: AC field activated and catalytic swimmers. The generality is highlighted by the collapse of all data in a single master curve, suggesting the applicability to further systems, both synthetic and biological.
\end{abstract}

\maketitle

\begin{figure*}[t!]
 \includegraphics[width=\textwidth]{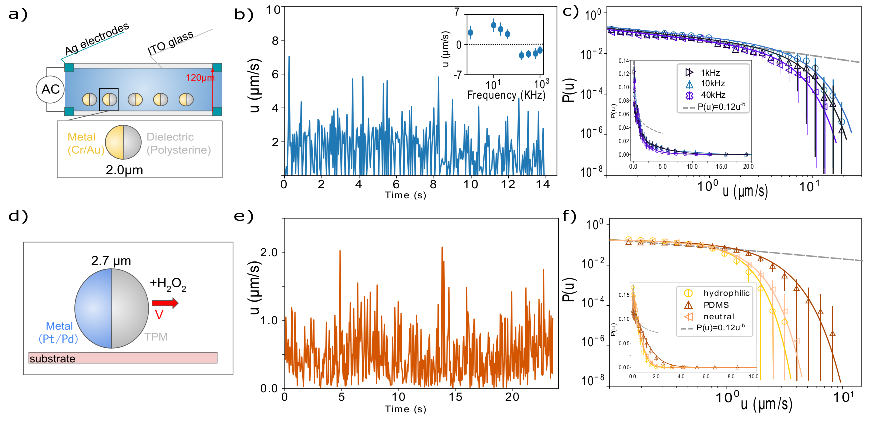}
\caption{\label{fig:1} {\bf Intermittent motion of colloidal swimmers activated in an electric field by ICEP/sDEP and by catalysis of $H_2O_2$} (a) Experimental setup for AC field activation of Janus spheres. (b) Typical velocity trace of a single particle, showing rapid fluctuations in velocity magnitude over time. Inset: Average velocity as a function of electric field frequency, showing a switch of sign at 120 KHz.  (c) Velocity distributions at three different frequencies in log-log representation (linear representation in inset). The velocities show a non-normal distribution, well fitted by a power-law with an exponential cut-off of the form $P(u)\propto u^{-b}e^{-cu}$. (d) Activation mechanism of the catalytic swimmers. (e) Velocity trace of a single catalytic particle, showing similar rapid fluctuations in velocity magnitude. (f) Corresponding velocity distributions for measured for three different surfaces with disparate contact angles, fitted with truncated power-law distributions similar as in (c).   }
\end{figure*}

Artificial swimmers have been the focus of a wide variety of research, with increasing interest in fields such as smart drug delivery~\cite{LeTC2019}, cargo transport in micro or nano-systems~\cite{Erez2022}, as well as the study of collective behavior in larger sytems~\cite{Harder2018} in active matter. There are also many examples of microswimmers in nature. Spanning across five orders of magnitude in size, the dynamics of active motion is used to understand the contagion of nanometer-sized viruses \cite{Norambuena2020}, the motility patterns of micrometer-sized bacteria \cite{Viola2021}, and can even be extended to larger organisms such as salmon swimming against currents~\cite{Newton2021}.

The common denominator of all these systems is their ability to draw energy from their environment to perform directed motion \cite{Bechinger2016,Elgeti2015}. A simple realization of an experimental model to study such active motion are Janus colloids, exhibiting two or more distinct surfaces, leading to a chemical gradient of solutes properties, resulting in a net propulsion of the particle. The mechanism with which the gradients are induced can vary from preferential catalytic reactions~\cite{Ebbens2018} to local binary solvent demixing induced by light absorption~\cite{Buttinoni_2012}, to AC field application~\cite{Boymelgreen2016}.

In all such examples, it is standard practice to report the average swimming velocity~\cite{Boymelgreen2016,Ketzetzi2020} or the controllability of the swimming direction~\cite{Buttinoni_2012} as a function of the driving force parameters such as fuel concentration, or frequency of applied AC fields. Intricate dynamics of different systems have also been reported, some system specific, and others more generalized~\cite{Ajdari2006}. Despite sharing multiple common traits, different active systems are usually being discussed separately. An exception to this is the study of bacteria swimming by Lisicki et. al. \cite{Lisicki2019}, in which the authors collected data of swimming micro-organisms and grouped them according to the swimming mechanism. However, the commonalities of many of these systems in terms of their physical hydrodynamic mechanism of active swimming have not been addressed, masking any underlying dynamics of active motion.

In this Letter, we study in detail two disparate colloidal swimmer systems: a  system activated by an AC field, and a system of catalytic swimmers, activated by the creation of a solute gradient via a catalytic reaction that propels the particle. We uncover generic intermittency in the active swimmer motion with velocity distributions exhibiting surprisingly robust power laws in both systems with active force-dependent exponential cut-off. We show that this generic form can be understood based on a simple model including a competition of diffusion and active motion setting the replenishment of the gradient driving the motion, and randomly modulated hydrodynamic interaction associated with wetting inhomogeneities of the nearby wall. The generality of these findings is demonstrated by collapse of all velocity distributions into a single master curve. These results, revealing an intriguing coupling mechanism in the generic intermittency of the active motion, are expected to apply not only to synthetic colloidal swimmers, but also more generally to biological swimmers driven by a gradient, such as in nutrition.

In the AC field experiments, we use Janus particles of radius $1.00 \pm 0.05 \mu$m made of polystyrene (PS) particles, coated with a thin layer of Cr and Au, $5\mathrm{nm}$ and $20\mathrm{nm}$, respectively, activated via two separate mechanisms based on the frequency of the applied field. In the low frequency regime ($f<10KHz$), the particles move due to induced charge electrophoresis (ICEP), whilst in the higher frequency regime, the dominating mechanism is self-dielectrophoresis (sDEP)\cite{Boymelgreen2016}. The particles are suspended at sufficiently low concentration in deionized water with 1mM MgSO$_4$, and the resulting colloidal sample is placed between two glass slides coated with a 95nm layer of conductive Indium-Tin-Oxide (ITO) with a 10nm layer of silica on top to prevent sticking \cite{Nishiguchi2018}. The glass slides are separated by a spacer of thickness $120\mu$m and connected to a function generator (GW Instek AFG-2005) as shown in Fig.~\ref{fig:1}(a), providing an AC electric field with constant amplitude of $V_{pp}=10$V and varying frequency in the range of 1 to 1000kHz. The resulting active motion of the particles is followed with a bright-field microscope (ZEISS Axio Vert A1) equipped with a 63x objective with NA=1.4, and recorded at a frame rate of 20s$^{-1}$. Particles are located with an accuracy of 25nm in the image plane using image analysis~\cite{TP}, and their instantaneous velocities determined from their displacement between frames by dividing by the fixed time interval $\Delta t = 0.05s$ between images.

AC-field activation allows tuning the magnitude of the active force with frequency of the alternating field. Previous work has extensively explored the non-monotonic relation of the active force magnitude to frequency~\cite{Boymelgreen2016}, passing through a cross-over frequency (COF) where the active force becomes 0 and switches sign. In agreement with these studies, we observe a similar average velocity-frequency relation, with velocity switching sign at $\sim $120kHz, as shown in Fig.~\ref{fig:1}b (inset). Following previous conventions, we define the sign of the velocity to be positive when the particle is leading the motion with the dielectric hemisphere, and negative vice versa. Surprisingly, when we resolve the motion with fine time resolution, we observe intermittency in the velocity even when the particle proceeds along straight paths, as shown for an individual particle trajectory in Fig.~\ref{fig:1}b, where high velocities are followed by rapid slowing down.

To study this intermittency in more detail, we plot velocity distributions acquired for many particles in Fig.~\ref{fig:1}c. We observe that at low velocities, the distribution follows a power law, $P(u)\propto u^{-b}$, with exponent $b=0.45 \pm 0.05$ (dashed line), while the high velocity regime is dominated by an exponential cut-off, as confirmed by fitting over two orders of magnitude in velocity and more than 4 orders of magnitude in probability (solid lines). We also checked that the finite tracking accuracy did not affect the distribution, see SI. These truncated power-law distributions remind of those in yielding amorphous materials~\cite{Barrat_review2018,Chikkadi2011,Denisov2016}, but are of different origin as discussed below.

To investigate the generality of this finding, we study a different active colloidal system of catalytic swimmers ~\cite{Ketzetzi2020,Shim2022}. Particles made from 3-(trimethoxysilyl) propyl methacrylate (TPM), with radius $R=1.35{\mu}$m (2.4\% size polydispersity), were half-coated with $4.9\pm 0.2$ nm Pt/Pd. These Janus spheres propel via preferential decomposition of H$_2$O$_2$ by the metallic hemisphere~\cite{Ke2010} (Fig.~\ref{fig:1}d). A constant active force is achieved by control of the fuel concentration, while different propelling speeds are realized through different surface wetting properties. Besides the different swimming mechanism, the particles also differ in their cap-core mass ratios: while the Cr/Au coated polystyrene particles consisted of a light core and a heavy cap (see SI), these Pt/Pd particles consist of a heavy TPM core and a much lighter cap, giving rise to different torques exerted by gravity. Measurements were taken with a 60x ELWD air objective (NA 0.7) on an inverted Nikon TI-E microscope at a frame rate of 19 s$^{-1}$ in the dark, typically within the hour after dispersing the colloids at dilute particle concentration $\sim 10^{-7}$ w/v) in deionized water containing 10\% H$_2$O$_2$.

Despite the different driving mechanism, and physical properties, we find very similar intermittency (Fig.~\ref{fig:1}e) and power-law velocity distributions with exponent $b^\prime = 0.46 \pm 0.06 $ for all three surfaces, shown in Fig.~\ref{fig:1}f, indistinguishable within error bars from $b$. The three different substrates have measured contact angles $\theta = 30^o\pm 3^o , 51^o\pm 3^o$ and $100^o\pm 3^o$, corresponding to a hydrophilic, a neutral, and a hydrophobic PDMS-coated glass surface. Previous measurements showed that the average velocity scales as $\langle u \rangle \propto (1+\mathrm{cos}(\theta))^{-3/2}$~\cite{Ketzetzi2020}. Here, we observe that the exponential cut-off shifts to the right accordingly with increasing contact angle, while the form of the distribution remains the same across the data.

\begin{figure}[t!]
\includegraphics[width=0.9\columnwidth]{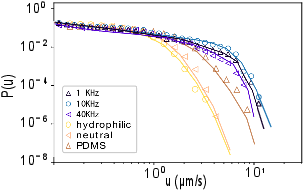}
\caption{\label{fig:2} {\bf Model predictions.} Velocity distributions of \com{Ac-field} and catalytic swimmers, for different frequencies and glass substrates respectively, in log-log representation for experimental (symbols) and model generated data (solid lines). The power law remains robust across both systems, different active forces, and different wetting angles.}
\end{figure}

The observation of power-law intermittency in two disparate systems suggests a general underlying driving mechanism. To describe it, we construct a simple model for the dynamics of self-propelling particles, where the driving force generated by the particle-induced gradient is balanced by a drag force experienced by the moving particle, which scales linearly with velocity at low Reynolds numbers~\cite{Anderson1989}. We assume that the driving force is constant as long as the particle maintains the local gradient, which decays exponentially within an interfacial layer, $\lambda << R$ \cite{Anderson1989}. At high velocities, however, the particle runs out of its gradient: due to the finite time it takes to generate the local gradient, we expect a fast moving particle to run ahead of its maximum gradient, thereby reducing its active force. For the AC system, the ionic gradient generated within the double layer decays within a characteristic length, the Debye length. Equivalently, for the catalytic swimmers, catalysis can occur when the H$_2$O$_2$ is in close proximity with the Pt cap, likewise defining a characteristic length at twice the size of the H$_2$O$_2$ molecule. 

To account for this effect, we describe the velocity-dependent active force by considering a first-order negative loop relationship, $F_a^\mathrm{eff}\propto F_a e^{-\frac{1}{u_\mathrm{th}}\frac{dx}{dt}} $, where $F_a$ is the maximum active force, $dx/dt$ is the instantaneous velocity expressed in differential form, and $u_\mathrm{th} \propto D/\lambda$ is a characteristic velocity threshold proportional to the ratio of the relevant diffusion constant $D$, and length, $\lambda$. The values of $u_\mathrm{th}$ for the AC-swimmers are calculated from the average diffusion coefficient ($1.07$ and $0.71 \times 10^{-3} \mu m^2/s$ for sulfate and magnesium ions, respectively~\cite{phreeqc}), and the smallest Debye length of the 1mM salt concentration ($1.69nm$ and $2.92$nm for the polysterene and Au hemispheres), giving $u_\mathrm{th} = 0.51\mu$m/s. For the catalytic system, we used the diffusion coefficient $1.75 \times 10^{-4}\mu m^2/s$ of $\mathrm{H_2O_2}$ and the characteristic adsorption length $\lambda = 5.95\times10^{-4}\mu m$, twice the size of an $H_2O_2$ molecule, yielding $u_\mathrm{th} = 0.34\mu$m/s.

The drag force that the particles experience close to the wall is modulated by a coefficient dependent on the slip length of the surface~\cite{Loussaief2015,Hocking1973}, which in turn depends on the contact angle $\theta$~\cite{Danov2000,Huang2008}. The contact angle is expected to vary randomly within certain limits along the substrate due to heterogeneities in the coating of the surface, as well as the particles' height above the surface \cite{Ketzetzi2020}(SI). We model this effect by a spatially varying drag coefficient, $c(x) \propto (1+\mathrm{cos}(\theta(x)))^{3/2}$, describing a random field of "obstacles" encountered by the particle, where hydrophilic patches will amplify the drag and hydrophobic patches will reduce it, creating an effective resistance to the motion \cite{Loussaief2015}. Together, the balance of active and drag forces leads to the equation of motion:
\begin{equation}
    0 = F_a e^{-\frac{1}{u_\mathrm{th}}\frac{dx}{dt}} - 6\pi \eta R c(x') \frac{dx}{dt} + \mathcal{F}(t)
    \label{eq1}
\end{equation}

\begin{figure*}[t!]
\includegraphics[width=1.0\textwidth]{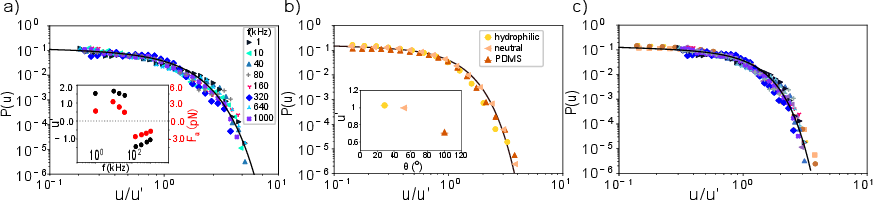}
\caption{\label{fig:3} {\bf Generic mechanism.} (a) Velocity distributions of the AC-field experiments, collapsed into a single master curve by scaling along the x-axis by $u^\prime$. The scaling parameter $u'$ and corresponding $F_a$ (inset) exhibit the same functional form as the average velocity in ~\ref{fig:1}b (inset). (b) Velocity distributions of the catalytic swimmers, collapsed by scaling along the x-axis, by the ratios of the contact angles $\theta$ (inset). (c) Collapse of all data for both systems demonstrates general underlying mechanism.}
\end{figure*}

where $\eta$ is the dynamic viscosity of the medium, $R$ the particle radius, and $\mathcal{F}(t)$ a random force term~\cite{Romanczuk_2012} that accounts for the random fluctuations, and we have set the inertial component (left-hand side) to $0$ since our system is overdamped. The velocity of fast moving particles will hence be limited by the diffusion-limited replenishment of the gradient ($1^{st}$ term), and the hydrodynamic drag ($2^{nd}$). Note that each of the two swimmer systems explores different values of the two terms on the right: the AC-field system the first term, varying the active force magnitude through frequency, and the catalytic swimmer system the second, varying the effective drag force through contact angle. Equation \ref{eq1} is an non-linear ordinary differential equation, which can be numerically solved for x (see SI).

We employ eq. 1 to simulate velocity traces in one dimension, using the calculated $u_\mathrm{th}$ value and the particle radii $R$ of the two systems, and the viscosity of water $\eta =1~$mPas. To implement small fluctuations in the wetting angle, we sample a narrow Gaussian distribution, centered at the measured value of $\theta$, after the particle has traversed a fixed length, resulting in a position-dependent $c(x)$. Hence, we differentiate $|x|$ with respect to time to obtain velocity and construct a distribution. The model uses a fixed temporal increment per step and a randomly generated 1-dimensional map of c(x) factors, where the time step is chosen small compared to the experimental resolution (0.5$\mu$s) so we can consider each velocity in that time step constant. 

Each system explores one of the two remaining parameters, $F_a$ and $c(x)$. To find $F_a$, we performed multiple runs, slowly increasing the active force, for both systems. The resulting velocity distributions where then compared with the experimental data. In the case of the AC-experiments, this was done with the same $c(x)$. For the catalytic swimmers, $c(x)$ was adapted to reflect the measured values of the wetting angle. Since the active force is defined by the concentration of $H_2O_2$, which was constant, the active force was selected such that the best fit was obtained for all three surfaces with respect to the experimental measurements.

Excellent fits are obtained with active forces $F_a = 2.5$, $3.5$, and $1.2$ pN at the different frequencies of $1$, $10$, and $40$kHz, respectively, for the AC-field system, as well as with $F_a = 0.5$pN and the different experimentally measured contact angles $\theta$ for the catalytic swimmers, as shown in Fig.~\ref{fig:2}. Small deviations arise at the higher velocities, which could be a result of the small amount of experimental data of high velocities. Hence, in both cases, the dynamics described by Eq.~\ref{eq1} are able to generate velocity distributions very similar to the experimental ones, capturing not only the power-law intermittency, but also the exponential cut-off, thus indicating the generality of the model.

To further highlight the generality of the underlying mechanism, we collapse the experimental data for all frequencies by re-scaling the velocity axis with $u'$. The collapsed data, together with the extracted scaling parameter $u'$ are shown in Fig.~\ref{fig:3}a. Indeed, $u'$ shows a very similar functional dependence as the average velocity in Fig.~\ref{fig:1}b inset, lending credence to our approach. We can similarly collapse the simulated data onto the same master curve; the extracted active force $F_a$ from the model shows a corresponding dependence on frequency, see inset.

A similar collapse is obtained for the catalytic swimmers for the three different surfaces, as shown in Fig.~\ref{fig:3}b. From the corresponding velocity scaling parameter, we can now determine the contact angles of the substrates; using the neutral surface as reference to determine the proportionality constant, we can then calculate the contact angles for the other two surfaces. We find values of  $29^o$ and $102^o$ for the hydrophilic and PDMS surface respectively, in good agreement with the experimentally measured contact angles of $26^o \pm 3^o$ and $100^o\pm 3^o$, supporting not only the relation proposed in~\cite{Ketzetzi2020}, but also highlighting the wetting-dependent drag force, as the effect is more pronounced at the high velocity regime.

We can also determine the three wetting angles directly from the model fits to the experimental velocity distributions. In this case, we allow the wetting angle $\theta$ to vary openly, whilst keeping $F_a$ bound within a narrow limit defined by previous runs utilizing the knowledge that $F_a$ has to be the same for all surfaces, as the fuel concentration is kept constant. Doing so, we find values of $29^o$, $49.5^o$ and $102^o$ for the hydrophilic, neutral, and PDMS surface from the best fit to the experimental data, which are in very good agreement with the experimentally measured values.

Finally, to show the robustness of our findings, we collapse all data of both systems in a single master curve as shown in Fig.~\ref{fig:3}c. The excellent collapse of all data for both active swimmer systems in a single curve, exploring the two different forces involved - active and friction force - highlights the general underlying mechanism. Specifically, the data suggest the same power-law exponent for both systems within error bars, and demonstrates the interplay between the hydrodynamic drag and velocity threshold as described by the model. The collapse demonstrates that changes in the magnitude of the active force as well as friction force via the contact angle affects the exponential cut-off, allowing the power law to extend to higher velocities for higher active forces or hydrophobic surfaces.

Our experiments and model provide a new perspective on the dynamics of active swimmers. Interesting intermittency arises due to i) finite replenishment rate of the fuel gradient and ii) varying surface drag fields due to the randomly fluctuating surface wetting properties. The resulting equation of motion balancing the active and hydrodynamic drag forces describes the experimentally observed intermittency and produces velocity distributions closely matching experimental measurements for two disparate systems.

The consistent power-law regime with robust exponent found for both systems suggests an underlying generic behavior. Interestingly, the two ingredients above are reminiscent of the basic local ingredients of depinning models, in which a local elastic driving force, set by the distance to a propagating front, together with a random field of obstacles leads to power-law velocity distributions with exponential cut-off. Yet, while the full description of depinning involves interaction of the different parts of the propagating front, this is not the case for colloidal swimmers at low density.

The observed large velocity spreads, as reported in \cite{Ebbens2018,Boymelgreen2016}, may be directly related to the effects described in this Letter, namely the active force decay with velocity, and the hydrodynamic coefficient, which cannot be described by a Gaussian distribution. These results suggest further general implications for active swimmers. The presence of a power-law regime indicates a velocity region where there is no negative feedback to the motion by the motion, apart from the hydrodynamic drag, followed by a sharp exponential cut-off, modulated by both the ability of the system to propel and the experienced drag. In fact, the observed intermittency is analogous to the intermittency expected for scallop-type swimmers in biological systems, as it contains all the necessary ingredients - non-reciprocal motion, broken symmetry, and a "relaxation" time before each propulsion action. While the scallop changes shape to achieve directional motion, active colloidal swimmers achieve the same result by changing the ionic or chemical concentration of their local surroundings.

\textit{Aknowledgements}. - N.O.K and P.S. acknowledge funding (Grant No. 680.91.124) from the Netherlands Organization for Scientific Research (NWO). D.J.K. gratefully acknowledges funding through an ERC starting grant RECONFMATTER (Grant Agreement No. 758383) from the European Research Council. The authors also acknowledge the usefull discussion with Sara Jabbari Farouji and the Mazi Jalaal.

\bibliography{Bibliography}

\begin{thebibliography}{28}%
\makeatletter
\providecommand \@ifxundefined [1]{%
 \@ifx{#1\undefined}
}%
\providecommand \@ifnum [1]{%
 \ifnum #1\expandafter \@firstoftwo
 \else \expandafter \@secondoftwo
 \fi
}%
\providecommand \@ifx [1]{%
 \ifx #1\expandafter \@firstoftwo
 \else \expandafter \@secondoftwo
 \fi
}%
\providecommand \natexlab [1]{#1}%
\providecommand \enquote  [1]{``#1''}%
\providecommand \bibnamefont  [1]{#1}%
\providecommand \bibfnamefont [1]{#1}%
\providecommand \citenamefont [1]{#1}%
\providecommand \href@noop [0]{\@secondoftwo}%
\providecommand \href [0]{\begingroup \@sanitize@url \@href}%
\providecommand \@href[1]{\@@startlink{#1}\@@href}%
\providecommand \@@href[1]{\endgroup#1\@@endlink}%
\providecommand \@sanitize@url [0]{\catcode `\\12\catcode `\$12\catcode `\&12\catcode `\#12\catcode `\^12\catcode `\_12\catcode `\%12\relax}%
\providecommand \@@startlink[1]{}%
\providecommand \@@endlink[0]{}%
\providecommand \url  [0]{\begingroup\@sanitize@url \@url }%
\providecommand \@url [1]{\endgroup\@href {#1}{\urlprefix }}%
\providecommand \urlprefix  [0]{URL }%
\providecommand \Eprint [0]{\href }%
\providecommand \doibase [0]{http://dx.doi.org/}%
\providecommand \selectlanguage [0]{\@gobble}%
\providecommand \bibinfo  [0]{\@secondoftwo}%
\providecommand \bibfield  [0]{\@secondoftwo}%
\providecommand \translation [1]{[#1]}%
\providecommand \BibitemOpen [0]{}%
\providecommand \bibitemStop [0]{}%
\providecommand \bibitemNoStop [0]{.\EOS\space}%
\providecommand \EOS [0]{\spacefactor3000\relax}%
\providecommand \BibitemShut  [1]{\csname bibitem#1\endcsname}%
\let\auto@bib@innerbib\@empty
\bibitem [{\citenamefont {Le}\ \emph {et~al.}(2019)\citenamefont {Le}, \citenamefont {Zhai}, \citenamefont {Chiu},\ and\ \citenamefont {Tran}}]{LeTC2019}%
  \BibitemOpen
  \bibfield  {author} {\bibinfo {author} {\bibfnamefont {T.}~\bibnamefont {Le}}, \bibinfo {author} {\bibfnamefont {J.}~\bibnamefont {Zhai}}, \bibinfo {author} {\bibfnamefont {W.}~\bibnamefont {Chiu}}, \ and\ \bibinfo {author} {\bibfnamefont {P.}~\bibnamefont {Tran}},\ }\href@noop {} {\bibfield  {journal} {\bibinfo  {journal} {Int J Nanomedicine}\ }\textbf {\bibinfo {volume} {14}},\ \bibinfo {pages} {6749} (\bibinfo {year} {2019})},\ \bibinfo {note} {dOI:10.2147/IJN.S169030}\BibitemShut {NoStop}%
\bibitem [{\citenamefont {Erez}\ \emph {et~al.}(2022)\citenamefont {Erez}, \citenamefont {Karshalev}, \citenamefont {Wu}, \citenamefont {Wang},\ and\ \citenamefont {Yossifon}}]{Erez2022}%
  \BibitemOpen
  \bibfield  {author} {\bibinfo {author} {\bibfnamefont {S.}~\bibnamefont {Erez}}, \bibinfo {author} {\bibfnamefont {E.}~\bibnamefont {Karshalev}}, \bibinfo {author} {\bibfnamefont {Y.}~\bibnamefont {Wu}}, \bibinfo {author} {\bibfnamefont {J.}~\bibnamefont {Wang}}, \ and\ \bibinfo {author} {\bibfnamefont {G.}~\bibnamefont {Yossifon}},\ }\href {\doibase https://doi.org/10.1002/smll.202101809} {\bibfield  {journal} {\bibinfo  {journal} {Small}\ }\textbf {\bibinfo {volume} {18}},\ \bibinfo {pages} {2101809} (\bibinfo {year} {2022})}\BibitemShut {NoStop}%
\bibitem [{\citenamefont {Harder}\ and\ \citenamefont {Cacciuto}(2018)}]{Harder2018}%
  \BibitemOpen
  \bibfield  {author} {\bibinfo {author} {\bibfnamefont {J.}~\bibnamefont {Harder}}\ and\ \bibinfo {author} {\bibfnamefont {A.}~\bibnamefont {Cacciuto}},\ }\href@noop {} {\bibfield  {journal} {\bibinfo  {journal} {Phys Rev E}\ }\textbf {\bibinfo {volume} {97}},\ \bibinfo {pages} {022603} (\bibinfo {year} {2018})}\BibitemShut {NoStop}%
\bibitem [{\citenamefont {Norambuena}\ \emph {et~al.}(2020)\citenamefont {Norambuena}, \citenamefont {Valencia},\ and\ \citenamefont {Guzm{\'a}n-Lastra}}]{Norambuena2020}%
  \BibitemOpen
  \bibfield  {author} {\bibinfo {author} {\bibfnamefont {A.}~\bibnamefont {Norambuena}}, \bibinfo {author} {\bibfnamefont {F.~J.}\ \bibnamefont {Valencia}}, \ and\ \bibinfo {author} {\bibfnamefont {F.}~\bibnamefont {Guzm{\'a}n-Lastra}},\ }\href {\doibase 10.1038/s41598-020-77860-y} {\bibfield  {journal} {\bibinfo  {journal} {Scientific Reports}\ }\textbf {\bibinfo {volume} {10}},\ \bibinfo {pages} {20845} (\bibinfo {year} {2020})}\BibitemShut {NoStop}%
\bibitem [{\citenamefont {Tokárová}\ \emph {et~al.}(2021)\citenamefont {Tokárová}, \citenamefont {Perumal}, \citenamefont {Nayak}, \citenamefont {Shum}, \citenamefont {Kašpar}, \citenamefont {Rajendran}, \citenamefont {Mohammadi}, \citenamefont {Tremblay}, \citenamefont {Gaffney}, \citenamefont {Martel}, \citenamefont {Nicolau},\ and\ \citenamefont {Nicolau}}]{Viola2021}%
  \BibitemOpen
  \bibfield  {author} {\bibinfo {author} {\bibfnamefont {V.}~\bibnamefont {Tokárová}}, \bibinfo {author} {\bibfnamefont {A.~S.}\ \bibnamefont {Perumal}}, \bibinfo {author} {\bibfnamefont {M.}~\bibnamefont {Nayak}}, \bibinfo {author} {\bibfnamefont {H.}~\bibnamefont {Shum}}, \bibinfo {author} {\bibfnamefont {O.}~\bibnamefont {Kašpar}}, \bibinfo {author} {\bibfnamefont {K.}~\bibnamefont {Rajendran}}, \bibinfo {author} {\bibfnamefont {M.}~\bibnamefont {Mohammadi}}, \bibinfo {author} {\bibfnamefont {C.}~\bibnamefont {Tremblay}}, \bibinfo {author} {\bibfnamefont {E.~A.}\ \bibnamefont {Gaffney}}, \bibinfo {author} {\bibfnamefont {S.}~\bibnamefont {Martel}}, \bibinfo {author} {\bibfnamefont {D.~V.}\ \bibnamefont {Nicolau}}, \ and\ \bibinfo {author} {\bibfnamefont {D.~V.}\ \bibnamefont {Nicolau}},\ }\href {\doibase 10.1073/pnas.2013925118} {\bibfield  {journal} {\bibinfo  {journal} {Proceedings of the National Academy of Sciences}\ }\textbf {\bibinfo {volume} {118}},\ \bibinfo {pages} {e2013925118} (\bibinfo
  {year} {2021})},\ \Eprint {http://arxiv.org/abs/https://www.pnas.org/doi/pdf/10.1073/pnas.2013925118} {https://www.pnas.org/doi/pdf/10.1073/pnas.2013925118} \BibitemShut {NoStop}%
\bibitem [{\citenamefont {Newton}\ \emph {et~al.}(2021)\citenamefont {Newton}, \citenamefont {Barry}, \citenamefont {Lothian}, \citenamefont {Main}, \citenamefont {Honkanen}, \citenamefont {Mckelvey}, \citenamefont {Thompson}, \citenamefont {Davies}, \citenamefont {Brockie}, \citenamefont {Stephen}, \citenamefont {Murray}, \citenamefont {Gardiner}, \citenamefont {Campbell}, \citenamefont {Stainer},\ and\ \citenamefont {Adams}}]{Newton2021}%
  \BibitemOpen
  \bibfield  {author} {\bibinfo {author} {\bibfnamefont {M.}~\bibnamefont {Newton}}, \bibinfo {author} {\bibfnamefont {J.}~\bibnamefont {Barry}}, \bibinfo {author} {\bibfnamefont {A.}~\bibnamefont {Lothian}}, \bibinfo {author} {\bibfnamefont {R.}~\bibnamefont {Main}}, \bibinfo {author} {\bibfnamefont {H.}~\bibnamefont {Honkanen}}, \bibinfo {author} {\bibfnamefont {S.}~\bibnamefont {Mckelvey}}, \bibinfo {author} {\bibfnamefont {P.}~\bibnamefont {Thompson}}, \bibinfo {author} {\bibfnamefont {I.}~\bibnamefont {Davies}}, \bibinfo {author} {\bibfnamefont {N.}~\bibnamefont {Brockie}}, \bibinfo {author} {\bibfnamefont {A.}~\bibnamefont {Stephen}}, \bibinfo {author} {\bibfnamefont {R.~O.}\ \bibnamefont {Murray}}, \bibinfo {author} {\bibfnamefont {R.}~\bibnamefont {Gardiner}}, \bibinfo {author} {\bibfnamefont {L.}~\bibnamefont {Campbell}}, \bibinfo {author} {\bibfnamefont {P.}~\bibnamefont {Stainer}}, \ and\ \bibinfo {author} {\bibfnamefont {C.}~\bibnamefont {Adams}},\ }\href {\doibase 10.1093/icesjms/fsab024}
  {\bibfield  {journal} {\bibinfo  {journal} {ICES Journal of Marine Science}\ }\textbf {\bibinfo {volume} {78}},\ \bibinfo {pages} {1730} (\bibinfo {year} {2021})},\ \Eprint {http://arxiv.org/abs/https://academic.oup.com/icesjms/article-pdf/78/5/1730/40323714/fsab024.pdf} {https://academic.oup.com/icesjms/article-pdf/78/5/1730/40323714/fsab024.pdf} \BibitemShut {NoStop}%
\bibitem [{\citenamefont {Bechinger}\ \emph {et~al.}(2016)\citenamefont {Bechinger}, \citenamefont {Di~Leonardo}, \citenamefont {L\"owen}, \citenamefont {Reichhardt}, \citenamefont {Volpe},\ and\ \citenamefont {Volpe}}]{Bechinger2016}%
  \BibitemOpen
  \bibfield  {author} {\bibinfo {author} {\bibfnamefont {C.}~\bibnamefont {Bechinger}}, \bibinfo {author} {\bibfnamefont {R.}~\bibnamefont {Di~Leonardo}}, \bibinfo {author} {\bibfnamefont {H.}~\bibnamefont {L\"owen}}, \bibinfo {author} {\bibfnamefont {C.}~\bibnamefont {Reichhardt}}, \bibinfo {author} {\bibfnamefont {G.}~\bibnamefont {Volpe}}, \ and\ \bibinfo {author} {\bibfnamefont {G.}~\bibnamefont {Volpe}},\ }\href {\doibase 10.1103/RevModPhys.88.045006} {\bibfield  {journal} {\bibinfo  {journal} {Rev. Mod. Phys.}\ }\textbf {\bibinfo {volume} {88}},\ \bibinfo {pages} {045006} (\bibinfo {year} {2016})}\BibitemShut {NoStop}%
\bibitem [{\citenamefont {Elgeti}\ \emph {et~al.}(2015)\citenamefont {Elgeti}, \citenamefont {Winkler},\ and\ \citenamefont {Gompper}}]{Elgeti2015}%
  \BibitemOpen
  \bibfield  {author} {\bibinfo {author} {\bibfnamefont {J.}~\bibnamefont {Elgeti}}, \bibinfo {author} {\bibfnamefont {R.~G.}\ \bibnamefont {Winkler}}, \ and\ \bibinfo {author} {\bibfnamefont {G.}~\bibnamefont {Gompper}},\ }\href {\doibase 10.1088/0034-4885/78/5/056601} {\bibfield  {journal} {\bibinfo  {journal} {Reports on Progress in Physics}\ }\textbf {\bibinfo {volume} {78}},\ \bibinfo {pages} {056601} (\bibinfo {year} {2015})}\BibitemShut {NoStop}%
\bibitem [{\citenamefont {Ebbens}\ and\ \citenamefont {Gregory}(2018)}]{Ebbens2018}%
  \BibitemOpen
  \bibfield  {author} {\bibinfo {author} {\bibfnamefont {S.~J.}\ \bibnamefont {Ebbens}}\ and\ \bibinfo {author} {\bibfnamefont {D.~A.}\ \bibnamefont {Gregory}},\ }\href {\doibase 10.1021/acs.accounts.8b00243} {\bibfield  {journal} {\bibinfo  {journal} {Accounts of Chemical Research}\ }\textbf {\bibinfo {volume} {51}},\ \bibinfo {pages} {1931} (\bibinfo {year} {2018})}\BibitemShut {NoStop}%
\bibitem [{\citenamefont {Buttinoni}\ \emph {et~al.}(2012)\citenamefont {Buttinoni}, \citenamefont {Volpe}, \citenamefont {Kümmel}, \citenamefont {Volpe},\ and\ \citenamefont {Bechinger}}]{Buttinoni_2012}%
  \BibitemOpen
  \bibfield  {author} {\bibinfo {author} {\bibfnamefont {I.}~\bibnamefont {Buttinoni}}, \bibinfo {author} {\bibfnamefont {G.}~\bibnamefont {Volpe}}, \bibinfo {author} {\bibfnamefont {F.}~\bibnamefont {Kümmel}}, \bibinfo {author} {\bibfnamefont {G.}~\bibnamefont {Volpe}}, \ and\ \bibinfo {author} {\bibfnamefont {C.}~\bibnamefont {Bechinger}},\ }\href {\doibase 10.1088/0953-8984/24/28/284129} {\bibfield  {journal} {\bibinfo  {journal} {Journal of Physics: Condensed Matter}\ }\textbf {\bibinfo {volume} {24}},\ \bibinfo {pages} {284129} (\bibinfo {year} {2012})}\BibitemShut {NoStop}%
\bibitem [{\citenamefont {Boymelgreen}\ \emph {et~al.}(2016)\citenamefont {Boymelgreen}, \citenamefont {Yossifon},\ and\ \citenamefont {Miloh}}]{Boymelgreen2016}%
  \BibitemOpen
  \bibfield  {author} {\bibinfo {author} {\bibfnamefont {A.}~\bibnamefont {Boymelgreen}}, \bibinfo {author} {\bibfnamefont {G.}~\bibnamefont {Yossifon}}, \ and\ \bibinfo {author} {\bibfnamefont {T.}~\bibnamefont {Miloh}},\ }\href {\doibase 10.1021/acs.langmuir.6b01758} {\bibfield  {journal} {\bibinfo  {journal} {Langmuir}\ }\textbf {\bibinfo {volume} {32}},\ \bibinfo {pages} {9540} (\bibinfo {year} {2016})}\BibitemShut {NoStop}%
\bibitem [{\citenamefont {Ketzetzi}\ \emph {et~al.}(2020)\citenamefont {Ketzetzi}, \citenamefont {de~Graaf}, \citenamefont {Doherty},\ and\ \citenamefont {Kraft}}]{Ketzetzi2020}%
  \BibitemOpen
  \bibfield  {author} {\bibinfo {author} {\bibfnamefont {S.}~\bibnamefont {Ketzetzi}}, \bibinfo {author} {\bibfnamefont {J.}~\bibnamefont {de~Graaf}}, \bibinfo {author} {\bibfnamefont {R.~P.}\ \bibnamefont {Doherty}}, \ and\ \bibinfo {author} {\bibfnamefont {D.~J.}\ \bibnamefont {Kraft}},\ }\href {\doibase 10.1103/PhysRevLett.124.048002} {\bibfield  {journal} {\bibinfo  {journal} {Phys. Rev. Lett.}\ }\textbf {\bibinfo {volume} {124}},\ \bibinfo {pages} {048002} (\bibinfo {year} {2020})}\BibitemShut {NoStop}%
\bibitem [{\citenamefont {Ajdari}\ and\ \citenamefont {Bocquet}(2006)}]{Ajdari2006}%
  \BibitemOpen
  \bibfield  {author} {\bibinfo {author} {\bibfnamefont {A.}~\bibnamefont {Ajdari}}\ and\ \bibinfo {author} {\bibfnamefont {L.}~\bibnamefont {Bocquet}},\ }\href {\doibase 10.1103/PhysRevLett.96.186102} {\bibfield  {journal} {\bibinfo  {journal} {Phys. Rev. Lett.}\ }\textbf {\bibinfo {volume} {96}},\ \bibinfo {pages} {186102} (\bibinfo {year} {2006})}\BibitemShut {NoStop}%
\bibitem [{\citenamefont {Lisicki}\ \emph {et~al.}(2019)\citenamefont {Lisicki}, \citenamefont {Velho~Rodrigues}, \citenamefont {Goldstein},\ and\ \citenamefont {Lauga}}]{Lisicki2019}%
  \BibitemOpen
  \bibfield  {author} {\bibinfo {author} {\bibfnamefont {M.}~\bibnamefont {Lisicki}}, \bibinfo {author} {\bibfnamefont {M.~F.}\ \bibnamefont {Velho~Rodrigues}}, \bibinfo {author} {\bibfnamefont {R.~E.}\ \bibnamefont {Goldstein}}, \ and\ \bibinfo {author} {\bibfnamefont {E.}~\bibnamefont {Lauga}},\ }\href@noop {} {\bibfield  {journal} {\bibinfo  {journal} {Elife}\ }\textbf {\bibinfo {volume} {8}} (\bibinfo {year} {2019})}\BibitemShut {NoStop}%
\bibitem [{\citenamefont {Nishiguchi}\ \emph {et~al.}(2018)\citenamefont {Nishiguchi}, \citenamefont {Iwasawa}, \citenamefont {Jiang},\ and\ \citenamefont {Sano}}]{Nishiguchi2018}%
  \BibitemOpen
  \bibfield  {author} {\bibinfo {author} {\bibfnamefont {D.}~\bibnamefont {Nishiguchi}}, \bibinfo {author} {\bibfnamefont {J.}~\bibnamefont {Iwasawa}}, \bibinfo {author} {\bibfnamefont {H.-R.}\ \bibnamefont {Jiang}}, \ and\ \bibinfo {author} {\bibfnamefont {M.}~\bibnamefont {Sano}},\ }\href {\doibase 10.1088/1367-2630/aa9b48} {\bibfield  {journal} {\bibinfo  {journal} {New Journal of Physics}\ }\textbf {\bibinfo {volume} {20}},\ \bibinfo {pages} {015002} (\bibinfo {year} {2018})}\BibitemShut {NoStop}%
\bibitem [{\citenamefont {Allan}\ and\ \citenamefont {Caswell}(2017)}]{TP}%
  \BibitemOpen
  \bibfield  {author} {\bibinfo {author} {\bibfnamefont {D.}~\bibnamefont {Allan}}\ and\ \bibinfo {author} {\bibfnamefont {T.}~\bibnamefont {Caswell}},\ }\href@noop {} {\enquote {\bibinfo {title} {Trackpy},}\ } (\bibinfo {year} {2017})\BibitemShut {NoStop}%
\bibitem [{\citenamefont {Nicolas}\ \emph {et~al.}(2018)\citenamefont {Nicolas}, \citenamefont {Ferrero}, \citenamefont {Martens},\ and\ \citenamefont {Barrat}}]{Barrat_review2018}%
  \BibitemOpen
  \bibfield  {author} {\bibinfo {author} {\bibfnamefont {A.}~\bibnamefont {Nicolas}}, \bibinfo {author} {\bibfnamefont {E.~E.}\ \bibnamefont {Ferrero}}, \bibinfo {author} {\bibfnamefont {K.}~\bibnamefont {Martens}}, \ and\ \bibinfo {author} {\bibfnamefont {J.-L.}\ \bibnamefont {Barrat}},\ }\href@noop {} {\bibfield  {journal} {\bibinfo  {journal} {Rev. Mod. Phys.}\ }\textbf {\bibinfo {volume} {90}},\ \bibinfo {pages} {045006} (\bibinfo {year} {2018})}\BibitemShut {NoStop}%
\bibitem [{\citenamefont {Chikkadi}\ \emph {et~al.}(2011)\citenamefont {Chikkadi}, \citenamefont {Wegdam}, \citenamefont {D.Bonn}, \citenamefont {Nienhuis},\ and\ \citenamefont {Schall}}]{Chikkadi2011}%
  \BibitemOpen
  \bibfield  {author} {\bibinfo {author} {\bibfnamefont {V.}~\bibnamefont {Chikkadi}}, \bibinfo {author} {\bibfnamefont {G.}~\bibnamefont {Wegdam}}, \bibinfo {author} {\bibnamefont {D.Bonn}}, \bibinfo {author} {\bibfnamefont {B.}~\bibnamefont {Nienhuis}}, \ and\ \bibinfo {author} {\bibfnamefont {P.}~\bibnamefont {Schall}},\ }\href@noop {} {\bibfield  {journal} {\bibinfo  {journal} {Phys. Rev. Lett.}\ }\textbf {\bibinfo {volume} {107}},\ \bibinfo {pages} {198303} (\bibinfo {year} {2011})}\BibitemShut {NoStop}%
\bibitem [{\citenamefont {Denisov}\ \emph {et~al.}(2016)\citenamefont {Denisov}, \citenamefont {Lorincz}, \citenamefont {Uhl}, \citenamefont {Dahmen},\ and\ \citenamefont {Schall}}]{Denisov2016}%
  \BibitemOpen
  \bibfield  {author} {\bibinfo {author} {\bibfnamefont {D.}~\bibnamefont {Denisov}}, \bibinfo {author} {\bibfnamefont {K.}~\bibnamefont {Lorincz}}, \bibinfo {author} {\bibfnamefont {J.}~\bibnamefont {Uhl}}, \bibinfo {author} {\bibfnamefont {K.}~\bibnamefont {Dahmen}}, \ and\ \bibinfo {author} {\bibfnamefont {P.}~\bibnamefont {Schall}},\ }\href@noop {} {\bibfield  {journal} {\bibinfo  {journal} {Nature Comm.}\ }\textbf {\bibinfo {volume} {7}},\ \bibinfo {pages} {10641} (\bibinfo {year} {2016})}\BibitemShut {NoStop}%
\bibitem [{\citenamefont {Shim}(2022)}]{Shim2022}%
  \BibitemOpen
  \bibfield  {author} {\bibinfo {author} {\bibfnamefont {S.}~\bibnamefont {Shim}},\ }\href {\doibase 10.1021/acs.chemrev.1c00571} {\bibfield  {journal} {\bibinfo  {journal} {Chemical Reviews}\ }\textbf {\bibinfo {volume} {122}},\ \bibinfo {pages} {6986} (\bibinfo {year} {2022})}\BibitemShut {NoStop}%
\bibitem [{\citenamefont {Ke}\ \emph {et~al.}(2010)\citenamefont {Ke}, \citenamefont {Ye}, \citenamefont {Carroll},\ and\ \citenamefont {Showalter}}]{Ke2010}%
  \BibitemOpen
  \bibfield  {author} {\bibinfo {author} {\bibfnamefont {H.}~\bibnamefont {Ke}}, \bibinfo {author} {\bibfnamefont {S.}~\bibnamefont {Ye}}, \bibinfo {author} {\bibfnamefont {R.~L.}\ \bibnamefont {Carroll}}, \ and\ \bibinfo {author} {\bibfnamefont {K.}~\bibnamefont {Showalter}},\ }\href {\doibase 10.1021/jp101193u} {\bibfield  {journal} {\bibinfo  {journal} {The Journal of Physical Chemistry A}\ }\textbf {\bibinfo {volume} {114}},\ \bibinfo {pages} {5462} (\bibinfo {year} {2010})}\BibitemShut {NoStop}%
\bibitem [{\citenamefont {Anderson}(1989)}]{Anderson1989}%
  \BibitemOpen
  \bibfield  {author} {\bibinfo {author} {\bibfnamefont {L.~J.}\ \bibnamefont {Anderson}},\ }\href@noop {} {\bibfield  {journal} {\bibinfo  {journal} {Ann. Rev. Fluid Mech.}\ }\textbf {\bibinfo {volume} {21}},\ \bibinfo {pages} {61} (\bibinfo {year} {1989})}\BibitemShut {NoStop}%
\bibitem [{\citenamefont {Parkhurst}(1995)}]{phreeqc}%
  \BibitemOpen
  \bibfield  {author} {\bibinfo {author} {\bibfnamefont {D.~L.}\ \bibnamefont {Parkhurst}},\ }\href {https://search.library.wisc.edu/catalog/999778194102121} {\emph {\bibinfo {title} {User{\&}{\#}39;s guide to PHREEQC : a computer program for speciation, reaction-path, advective-transport, and inverse geochemical calculations}}}\ (\bibinfo  {publisher} {Lakewood, Colo. : U.S. Dept. of the Interior, U.S. Geological Survey ; Denver, CO : Earth Science Information Center, Open-File Reports Section [distributor], 1995.},\ \bibinfo {year} {1995})\ \bibinfo {note} {includes bibliographical references (pages 128-129).}\BibitemShut {Stop}%
\bibitem [{\citenamefont {Loussaief}\ \emph {et~al.}(2015)\citenamefont {Loussaief}, \citenamefont {Pasol},\ and\ \citenamefont {Feuillebois}}]{Loussaief2015}%
  \BibitemOpen
  \bibfield  {author} {\bibinfo {author} {\bibfnamefont {H.}~\bibnamefont {Loussaief}}, \bibinfo {author} {\bibfnamefont {L.}~\bibnamefont {Pasol}}, \ and\ \bibinfo {author} {\bibfnamefont {F.}~\bibnamefont {Feuillebois}},\ }\href {\doibase 10.1093/qjmam/hbv001} {\bibfield  {journal} {\bibinfo  {journal} {The Quarterly Journal of Mechanics and Applied Mathematics}\ }\textbf {\bibinfo {volume} {68}},\ \bibinfo {pages} {115} (\bibinfo {year} {2015})},\ \Eprint {http://arxiv.org/abs/https://academic.oup.com/qjmam/article-pdf/68/2/115/6867473/hbv001.pdf} {https://academic.oup.com/qjmam/article-pdf/68/2/115/6867473/hbv001.pdf} \BibitemShut {NoStop}%
\bibitem [{\citenamefont {Hocking}(1973)}]{Hocking1973}%
  \BibitemOpen
  \bibfield  {author} {\bibinfo {author} {\bibfnamefont {L.~M.}\ \bibnamefont {Hocking}},\ }\href {\doibase 10.1007/BF01535282} {\bibfield  {journal} {\bibinfo  {journal} {Journal of Engineering Mathematics}\ }\textbf {\bibinfo {volume} {7}},\ \bibinfo {pages} {207} (\bibinfo {year} {1973})}\BibitemShut {NoStop}%
\bibitem [{\citenamefont {Danov}\ \emph {et~al.}(2000)\citenamefont {Danov}, \citenamefont {Dimova},\ and\ \citenamefont {Pouligny}}]{Danov2000}%
  \BibitemOpen
  \bibfield  {author} {\bibinfo {author} {\bibfnamefont {K.~D.}\ \bibnamefont {Danov}}, \bibinfo {author} {\bibfnamefont {R.}~\bibnamefont {Dimova}}, \ and\ \bibinfo {author} {\bibfnamefont {B.}~\bibnamefont {Pouligny}},\ }\href {\doibase 10.1063/1.1289692} {\bibfield  {journal} {\bibinfo  {journal} {Physics of Fluids}\ }\textbf {\bibinfo {volume} {12}},\ \bibinfo {pages} {2711} (\bibinfo {year} {2000})},\ \Eprint {http://arxiv.org/abs/https://aip.scitation.org/doi/pdf/10.1063/1.1289692} {https://aip.scitation.org/doi/pdf/10.1063/1.1289692} \BibitemShut {NoStop}%
\bibitem [{\citenamefont {Huang}\ \emph {et~al.}(2008)\citenamefont {Huang}, \citenamefont {Sendner}, \citenamefont {Horinek}, \citenamefont {Netz},\ and\ \citenamefont {Bocquet}}]{Huang2008}%
  \BibitemOpen
  \bibfield  {author} {\bibinfo {author} {\bibfnamefont {D.~M.}\ \bibnamefont {Huang}}, \bibinfo {author} {\bibfnamefont {C.}~\bibnamefont {Sendner}}, \bibinfo {author} {\bibfnamefont {D.}~\bibnamefont {Horinek}}, \bibinfo {author} {\bibfnamefont {R.~R.}\ \bibnamefont {Netz}}, \ and\ \bibinfo {author} {\bibfnamefont {L.}~\bibnamefont {Bocquet}},\ }\href {\doibase 10.1103/PhysRevLett.101.226101} {\bibfield  {journal} {\bibinfo  {journal} {Phys. Rev. Lett.}\ }\textbf {\bibinfo {volume} {101}},\ \bibinfo {pages} {226101} (\bibinfo {year} {2008})}\BibitemShut {NoStop}%
\bibitem [{\citenamefont {Romanczuk}\ \emph {et~al.}(2012)\citenamefont {Romanczuk}, \citenamefont {Bär}, \citenamefont {Ebeling}, \citenamefont {Lindner},\ and\ \citenamefont {Schimansky-Geier}}]{Romanczuk_2012}%
  \BibitemOpen
  \bibfield  {author} {\bibinfo {author} {\bibfnamefont {P.}~\bibnamefont {Romanczuk}}, \bibinfo {author} {\bibfnamefont {M.}~\bibnamefont {Bär}}, \bibinfo {author} {\bibfnamefont {W.}~\bibnamefont {Ebeling}}, \bibinfo {author} {\bibfnamefont {B.}~\bibnamefont {Lindner}}, \ and\ \bibinfo {author} {\bibfnamefont {L.}~\bibnamefont {Schimansky-Geier}},\ }\href {\doibase 10.1140/epjst/e2012-01529-y} {\bibfield  {journal} {\bibinfo  {journal} {The European Physical Journal Special Topics}\ }\textbf {\bibinfo {volume} {202}},\ \bibinfo {pages} {1} (\bibinfo {year} {2012})}\BibitemShut {NoStop}%
\end{thebibliography}%

\end{document}